# MERGERS AND THE FORMATION OF DISK GALAXIES IN HIERARCHICALLY CLUSTERING UNIVERSES

M. STEINMETZ

*Max–Planck–Institut für Astrophysik*

*Postfach 1523, D–85740 Garching, Germany*

## 1. Introduction

It is commonly accepted, that galaxies acquire their angular momentum from tidal torques due to surrounding matter (Hoyle 1949). Most of the angular momentum is gained during the linear phase of the collapse (White 1984). After the turnaround point is reached, any further gain of angular momentum is strongly suppressed because it is increasingly difficult for the tidal field to act on the shrinking radius of the forming dark halo.

Due to its high cooling capabilities, the gas within the halo, which at turnaround possesses an angular momentum distribution similar to the dark halo, is virtually isothermal. In the standard picture, as it was essentially drawn by Fall & Efstathiou (1980), this gas cannot be supported by pressure and collapses into a thin, rotationally supported disk. The scale length of the disk is determined by the spin parameter $\lambda$ of the dark matter halo and the dark matter to baryon ratio $f_b$ within the halo. For reasonable values for the spin parameter ($\lambda \approx 0.07$, Barnes & Efstathiou 1987, Steinmetz & Bartelmann 1995) and $f_b \gtrsim 10$, the expected scale length ($\approx 3\,\mathrm{kpc}$) of a Milky–Way sized galaxy is in good agreement with observations.

This picture, however, only holds if no angular momentum is transported, i.e. the galaxy collapses axisymmetrically. If, however, galaxies form hierarchically, they are successively built up by merging of smaller structures (see, e.g., Lacey & Cole 1993). But mergers are typically identified with the sites where ellipticals are formed, but not spirals. Even more, numerical simulations of merging spirals (see e.g. Barnes, this volume) show that a huge amount of the angular momentum is transported to the dark halo and a very slowly rotating object is left. In a recent study, Navarro, Frenk & White (1995) found, that in simulations which start from cosmological initial conditions the specific angular momentum of the gas in the central gas knot is only 20% of that of the dark halo. This implies that these disks should possess a much smaller scale length than observed. However, the numerical resolution of these simulations was not sufficient to investigate the inner structure of the central gas knots. Especially it is an open question whether these dense knot of gas represents disks at all.

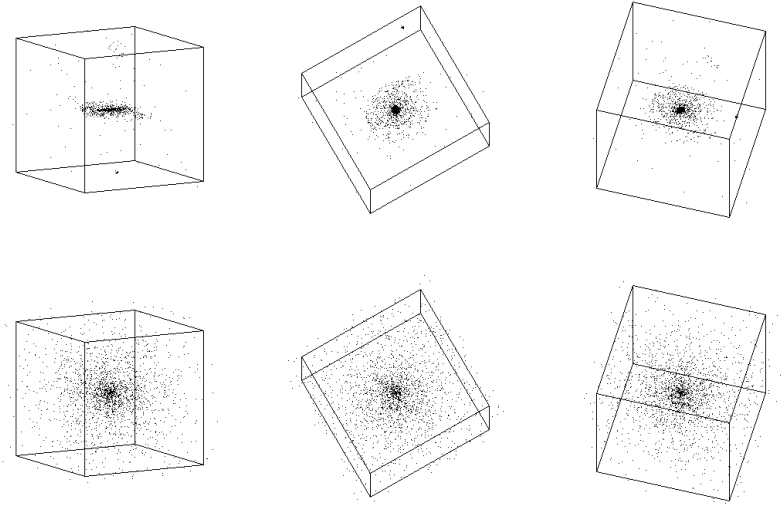

Figure 1. Three dimensional view on the distribution of gas (top) and dark matter (bottom) in a Milky Way sized halo ($v_c = 200\,\mathrm{km/sec}$) for three different projections. The box has a side length of 50 kpc.

## 2. The structure of disk galaxies in gasdynamical simulations

In order to analyze the distribution of gas near the centers of galaxies more detailed, we performed numerical simulations which possesses a roughly 5 times higher resolution than others have done so far. Initial conditions and boundaries were treated similarly to Navarro *et al.* (1995): Based on a coarsely grained, large scale N–body simulation (CDM, $\Omega = 1$, $\sigma_8 = 0.67$, $H_0 = 50$ km/sec/Mpc), individual haloes are picked out. Every halo is re-calculated in a high–resolution run. Gas dynamical effects are included using smoothed particle hydrodynamics (SPH). A very nice impression how structures built up and how mergers can cause a huge amount of angular momentum transport can be visualized by a movie (Steinmetz 1996).

Figure 1 shows the distribution of gas and dark matter for a typical halo formed in one of the simulations. At first glance, it seems to be quite similar to observed galaxies: the model galaxy consists of an extended spheroidal halo with a radius of about 300 kpc. The gas is much more centrally condensed and has settled in a thin disk with a diameter of about 50 kpc. A more careful analysis, however, shows that more than 70% of the gas can be found in the central gas clump within a radius of 2 kpc, the gravitational softening length. This gas has almost no angular momentum. The specific angular momentum of the disk, however, is similar to that of the dark halo. Consequently, the specific angular momentum of all the gas, within the disk and the dense gas knot, is a factor of about four smaller than that of the halo, in agreement with the simulations of Navarro *et al.* (1995). Compared to observations, the numerical models predict the formation of disks, which have the right size, but far too much mass is acquired by the central "bulge".

A deeper look at the formation history of the galaxy shows that the disk is mainly formed at low redshift ($z < 1$) by accretion of diffusely distributed gas. The dense gas knot, however, forms preferentially at higher redshifts ($z > 2$) by several merging events. In order to solve the angular momentum problem, one has to take care, that more gas is diffusely accreted.

## 3. Summary and discussion

Up to now, there is no self–consistent numerical simulation which avoids this angular momentum problem. However a variety of different processes are currently under discussion, which might be able to solve this problem. In my personal view, the first three can already be ruled out.

1. *Numerical imperfections.* The numerical viscosity which is usually used in SPH to describe shock waves (Gingold & Monaghan 1983) do not vanish in pure shear flows and, therefore, may cause an artificial angular momentum transport. We used a modified formulation which was shown not to exhibit a significant viscous angular momentum transport over several Hubble times (for details see Steinmetz 1995a and references therein).
2. *Changing the cosmological parameters* $\Omega_0$, $\Lambda_0$. This probably has rather little influence: Although the merging rate in the near past can be changed a lot, the merging history, expressed in terms of number of mergers and mass distribution of progenitors, is quite similar (Lacey & Cole, 1993).
3. *Photoionisation by a UV background field.* A UV background with a strength as inferred from quasar observations at high redshifts ($z = 1\ldots 4$) may suppress the formation of small structures ($v_c \lesssim 50$ km/sec, Efstathiou 1992) and the gas might fall in more diffusely. Therefore, the angular momentum transport to the halo might be reduced. We have performed numerical simulations including a UV background field.

   It turns out that the formation of the dense central clump is only marginally affected: Most of the gas in the clump collapses at high redshift, i.e. at high densities, where the influence of a UV background is small (Steinmetz 1995b). However the background field is able to prevent cooling of the late and diffusely infalling gas, which possesses a low density. Therefore, the formation of the disk found in the simulations without a UV background is almost completely suppressed.
4. *Feedback processes.* The most likely solution to the angular momentum problem is related to star formation and supernovae feedback. As a first approach we performed a simulation, where regions of rapidly cooling and convergent gas flow are assumed to form stars. In this model supernovae increase the thermal energy of the surrounding gas. However, due to the very high cooling capability of this gas, most of the energy is radiated away and so the efficiency of the supernovae feedback is rather low. As a result, the formation of small lumps of gas is not prevented. The main influence of star formation is to transform a dense knot of gas into a slightly more diffuse lump of stars, but the extensive transport of angular momentum to the dark halo is not overcome. It is conceivable that momentum input due to supernovae might have a much stronger effect (Navarro & White 1993): In contrast to thermal energy, kinetic energy cannot immediately be radiated away resulting in a much higher efficiency of the supernova feedback.

PUSTIL'NIK: What mass fraction is left in you models in gas clumps which are not in dark matter haloes:
STEINMETZ: Almost all gas clumps are hosted by a dark halo. Only those objects which are going to be accreted by a more massive galaxy have sometimes tidally stripped their dark halo. It may, however, well be a numerical artifact. These objects are very small in mass and are represented by a few dozens of particles. In the case of a higher numerical resolution, the maximum possible phase space increases and the dark halo may (partially) survive the encounter.

VÖLK: If it is just a question of angular momentum *redistribution*, would not the easiest way to achieve this be an outflow (with magnetic field)?

STEINMETZ: As I already mentioned I agree that the solution to the angular momentum problem has most likely to do with feedback processes (which also may produce outflow of gas or galactic winds). Note, however, that the main redistribution is transport of angular momentum from the gas to the dark matter.

RENZINI: Can the code make both, spirals *and* ellipticals.

STEINMETZ: In the simulations I showed in the movie, star formation was not included. Therefore, galaxies always end up as spiral like objects. However, you have seen a halo which was involved in a very violent merging event which was even able to flip the spin of the galaxy. In a simulation including star formation such a merger would most probable form an elliptical galaxy.